%% file: typeinst.tex
\documentclass[runningheads,a4paper]{llncs}

\usepackage{amssymb}
\usepackage{graphicx}

\usepackage{url}

\urldef{\mailsa}\path|{nqminh, giovanni, goran_topic, aizawa}@nii.ac.jp|    
\newcommand{\keywords}[1]{\par\addvspace\baselineskip
\noindent\keywordname\enspace\ignorespaces#1}

\usepackage{array}
\newcolumntype{L}[1]{>{\raggedright\let\newline\\\arraybackslash\hspace{0pt}}m{#1}}
\newcolumntype{C}[1]{>{\centering\let\newline\\\arraybackslash\hspace{0pt}}m{#1}}
\newcolumntype{R}[1]{>{\raggedleft\let\newline\\\arraybackslash\hspace{0pt}}m{#1}}

\begin{document}

\mainmatter  % start of an individual contribution

\title{A hybrid approach for semantic enrichment of MathML mathematical expressions}

\titlerunning{Semantic enrichment of mathematical expressions}

\author{Minh-Quoc Nghiem$^{~1}$, Giovanni Yoko Kristianto$^{~2}$,\\ Goran Topi\'{c}$^{~3}$, Akiko Aizawa$^{~2,3}$}
\authorrunning{M.Q. Nghiem, G.Y. Kristianto, G. Topi\'{c}, A. Aizawa}

\institute{$^{1~}$The Graduate University for Advanced Studies,\\
$^{2~}$The University of Tokyo,\\
$^{3~}$National Institute of Informatics,\\
Tokyo, Japan\\
\mailsa\\
}

\toctitle{Lecture Notes in Computer Science}
\tocauthor{Authors' Instructions}
\maketitle

\begin{abstract}
In this paper, we present a new approach to the semantic enrichment of mathematical expression problem.
Our approach is a combination of statistical machine translation and disambiguation which makes use of surrounding text of the mathematical expressions.
We first use Support Vector Machine classifier to disambiguate mathematical terms using both their presentation form and surrounding text.
We then use the disambiguation result to enhance the semantic enrichment of a statistical-machine-translation-based system.
Experimental results show that our system archives improvements over prior systems.
\end{abstract}

\input{1-intro}

\input{2-related}

\input{3-approach}

\input{4-eva}

\input{5-con}

\end{document}

%% file: 1-intro.tex
\section{Introduction}

% Semantic Enrichment of mathematical expression
The semantic enrichment of mathematical documents is among the most significant areas of math-aware technologies.
It is the process of associating semantic tags, usually concepts, with mathematical expressions.
We use MathML~\cite{MathML} Presentation and Content Markup to represent mathematical expressions and their meaning.
The semantic enrichment task then becomes the task of generating Content MathML outputs from Presentation MathML expressions.
It is an important technology towards fulfilling the dream of global digital mathematical library (DML).

% Challenges (semantic enrichment and disambiguation)
The semantic enrichment of mathematical expression is a challenging task.
Mathematical notations are ambiguous, context-dependent, and vary from community to community.
Given a Presentation MathML element, there are many potential mappings to its Content MathML element.
For example, the token $\delta$ can be mapped to $KroneckerDelta$, $DiracDelta$, $DiscreteDelta$, or $\delta$.
By correctly disambiguating these token elements, we can get a more accurate semantic enrichment system.

% Previous approach (semantic enrichment and disambiguation)
Disambiguation of mathematical elements is an important component in the semantic enrichment system.
Basic methods for dealing with ambiguities so far were either rule-based~\cite{STeX} or statistics-based~\cite{JSAI}.
The rule-based approach is of course generally not able to derive meaning from arbitrary Presentation MathML expressions.
The statistics-based approach resolves ambiguities based on the probabilities, and thus gets better results than the rule-based system.
In this paper, we enhance the statistics-based approach by combining it with a disambiguation component.

% Limitation of previous approach
So far, there has been limited discussion about the contribution of surrounding text to mathematical element disambiguation problem.
It is becoming increasingly difficult to ignore the surrounding text of mathematical expressions.
For example, the token $\delta$ can be mapped to $KroneckerDelta$ if its surrounding text contains the word `Kronecker delta'.
It is difficult to disambiguate using only the presentation of mathematical expression.
The combination of mathematical expression itself and its surrounding text can lead to improvements in disambiguation process.

% Problem to solve
The aim of this paper is to examine and solve the ambiguity when mapping Presentation MathML elements to their Content elements.
This paper also attempts to find the contribution of surrounding text to mathematical element disambiguation problem.
We use a Support Vector Machine (SVM) learning model for MathML Presentation token element (mi) disambiguation.
Both presentation of mathematical expression and its surrounding text are encoded in a feature vector used in SVM.
We evaluate the efficacy of the system by incorporating it into an SMT-based semantic enrichment system.

% Our contributions
We formulate the problem as follows: given a Presentation MathML expression and its surrounding text, can we interpret its Content MathML expression?
This paper provides contributions in three main areas of mathematical semantic enrichment problem.
First, we show that combination of a disambiguation component and the SMT-based system improves the system's performance.
Second, we show that the text surrounding the mathematical expressions contributes to the disambiguation process.
Third, we show that the name of the category that a mathematical expression belongs to is the most important text feature for disambiguation.

% Structure of paper
The remainder of this paper is organized as follows.
Sections 2 provides a brief overview of the background and related work on semantic enrichment of mathematical expressions. Section 3 presents our method.
Section 4 describes the experimental setup and results.
Section 5 concludes the paper and points to avenues for future work.

%% file: 2-related.tex
\section{Related Work}

% MathML
MathML~\cite{MathML} is the best-known open markup format for representing mathematical formulas.
It is recommended by the W3C Math Working Group as a standard to represent mathematical expressions.
MathML is an application of XML for describing mathematical notations and encoding mathematical content within a text format.
MathML has two types of encoding: Content MathML addresses the meaning of formulas; and Presentation MathML addresses the display of formulas.
We use MathML Presentation Markup to display mathematical expressions and MathML Content Markup to convey mathematical meaning.

% Mathematica
Most major computer algebra systems, such as Mathematica~\cite{Mathematica} and Maple~\cite{Maple}, are capable of importing and exporting MathML of both formats.
These importing and exporting functions enable the conversion from Presentation to Content MathML.
Importing, of course, depends on the interpretation of each computer algebra systems engine.

% SnuggleTeX
There is a project called SnuggleTeX~\cite{STeX}, which addresses the semantic interpretation of mathematical expressions.
The project provides a direct way to generate Content MathML from Presentation MathML based on manually encoded rules.
The current version at the time of writing this paper supports operators that are the same as ASCIIMathML~\cite{AMM}.
For example, it uses the ASCII string ``\textbackslash $ in $'' instead of the symbol ``$ \in $''.
One major drawback of this approach is that it always makes the same interpretation for the same Presentation MathML element.

% JSAI
A recent study by Nghiem et al.~\cite{JSAI} also addressed the semantic interpretation of mathematical expressions.
This study applied a method based on statistical machine translation to extract translation rules automatically.
This approach contrasted with previous research, which tended to rely on manually encoded rules.
This study also introduced segmentation rules used to segment mathematical expressions.
Combining segmentation rules and translation rules strengthened the translation system and the best system achieved 20.89\% error rate.
The shortcoming of this approach is that it did not make use of text information surrounding mathematical expressions.

% symbolic expression classify
Wolska et al.~\cite{DIS,DIS2} presented a knowledge-poor method of finding a denotation of simple symbolic expressions in mathematical discourse.
The system used statistical co-occurrence measures to classify a simple symbolic expression into one of seven predefined concepts.
They showed that the lexical information from the linguistic context immediately surrounding the expression improved the results.
The lexical information from the larger document context also contributed to the best interpretation results.
This approach had been evaluated on a gold standard manually annotated by experts, achieving 66\% precision.

%% file: 3-approach.tex
\section{Our Approach}

\begin{figure*}[htbt]
\centering
\includegraphics[scale=0.45]{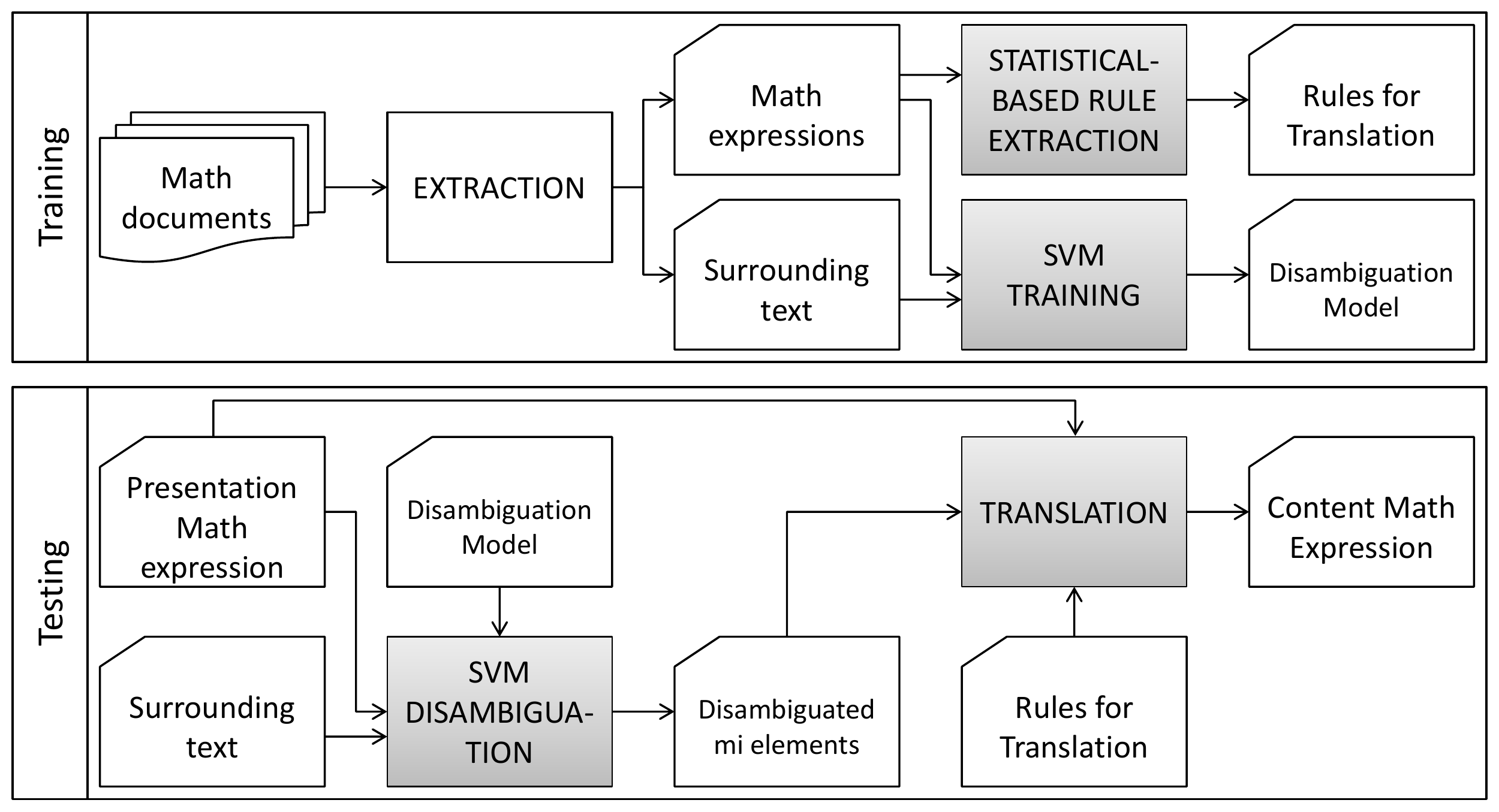}
\caption{System Framework}
\label{fig:sys}
\end{figure*}

% Framework
The system has two phases, a training phase and a running phase, and consists of three main modules.
\begin{itemize}
\item Statistical-based rule extraction: Extracts rules for translation, given the training data. We establish two types of rules: segmentation rules and translation rules. Each rule is associated with its probability.
\item SVM-based disambiguation: An SVM training algorithm builds a model that assigns to identifiers ($mi$) their correct content. Features are extracted from both the presentation of mathematical expressions and their surrounding text.
\item Translation: The input of this module includes a Presentation MathML expression, a set of rules for translation, and the output from the disambiguation module. This module translates Presentation into Content MathML expression.
\end{itemize}
Figure \ref{fig:sys} shows the system framework.

% SMT-based rule extraction
\subsection{Statistical-based rule extraction}
The rules for translation were extracted according to the procedure used by Nghiem et al.~\cite{JSAI}.
Given a set of training mathematical expressions in MathML parallel markup, we extracted two types of rules: segmentation rules and translation rules.
Translation rules are used to translate (sub)trees of Presentation MathML markup to (sub)trees of Content MathML markup.
Segmentation rules are used to combine and reorder the (sub)trees to form a complete tree.
The output of this module is a set of segmentation and translation rules, each rule is associated with its probability.

% Difference
%There are a number of important differences between X and Y. 

% SVM classification
\subsection{SVM disambiguation}
An $mi$ token element in MathML presentation markup can be translated into many different elements in MathML content markup.
In this paper, we assumed that one $mi$ element can be translated into one of a limited predefined set of Content elements.
Given an $mi$ element, we use an SVM training algorithm to build a model that assigns to its correct Content element.
When translating, each of the Presentation $mi$ elements will be disambiguated before generating Content MathML expressions.
The accuracy of the SVM disambiguation is a crucial preprocessing step for a high-quality MathML Presentation to Content translation.

% Generate training and testing data
We used the alignment output of GIZA++\footnote{https://code.google.com/p/giza-pp/}~\cite{GIZA} to generate training and testing data for the disambiguation problem.
Given a training data consists of several parallel markup expressions, we used GIZA++ to align the Presentation terms to the Content terms.
From this alignment results, we extract pairs of Presentation $mi$ elements and their associated Content elements.
Only $mi$ elements that have ambiguities in their translation are kept to generate training and testing data.
Table shows \ref{tab:ambi} the examples of Presentation $mi$ elements and their associated Content elements.

\begin{table}[htb]
\centering
\caption{Presentation $mi$ elements and their associated Content elements}
\begin{tabular}{|L{3.2cm}|L{4.8cm}|}
\hline
\centering\textbf{Presentation elements} 	& \centering\textbf{Content elements} \tabularnewline \hline
$<$mi$>\sigma<$/mi$>$ 	& $<$ci$>$Weierstrass Sigma$<$/ci$>$ \\ \cline{2-2}
						& $<$ci$>$Divisor Sigma$<$/ci$>$ \\ \cline{2-2}
						& $<$ci$>\sigma<$/ci$>$ \\ \hline
$<$mi$>\mu<$/mi$>$ 	& $<$ci$>$MoebiusMu$<$/ci$>$ \\ \cline{2-2}
						& $<$ci$>\mu<$/ci$>$ \\ \hline
$<$mi$>$H$<$/mi$>$ 		& $<$ci$>$StruveH$<$/ci$>$ \\ \cline{2-2}
						& $<$ci$>$Harmonic Number$<$/ci$>$ \\ \cline{2-2}
						& $<$ci$>$Hankel H1$<$/ci$>$ \\ \cline{2-2}
						& $<$ci$>$Hankel H2$<$/ci$>$ \\ \cline{2-2}
						& $<$ci$>$Hermite H2$<$/ci$>$ \\ \cline{2-2}
						& $<$ci$>$H$<$/ci$>$ \\ \hline
$<$mi$>$y$<$/mi$>$ 		& $<$ci$>$Bessel Y Zero$<$/ci$>$ \\ \cline{2-2}
						& $<$ci$>$Spherical Bessel Y$<$/ci$>$ \\ \cline{2-2}
						& $<$ci$>$y$<$/ci$>$ \\ \hline
\end{tabular}
\label{tab:ambi}
\end{table}

%Create positive and negative examples
For each mathematical expression, an $mi$ element has only one correct translation.
In other mathematical expressions, the same $mi$ element might have another correct translation.
Assume that an $mi$ element $e$ has $n$ ways of translating from Presentation into Content MathML.
For each mathematical expression, we create one positive instance by combining $e$ and its correct translation.
We also create $n-1$ negative instances by combining $e$ and its incorrect translations.

% Classify mathematical terms
The features used in the SVM disambiguation may be divided into two main groups: Presentation MathML features and surrounding text features.
Presentation MathML features are extracted from the Presentation MathML markup of the mathematical expression.
Surrounding text features are extracted from the text surrounding the mathematical expression.
The category which the mathematical expression belongs to is also used.
Table \ref{tab:fea} shows the features we used for classification.

\begin{table}[htb]
\centering
\caption{Features used for classification}
\begin{tabular}{|C{2cm}|L{3cm}|L{5cm}|}
\hline
\multicolumn{2}{|c|}{\centering\textbf{Feature}} 	& \centering\textbf{Description} \tabularnewline \hline
Presentation MathML 		& Only child			& Is it the only child of its parent node 	\\ \cline {2 -3}
feature						& Preceded by mo 		& Is it preceded by an $<$mo$>$ node 		\\ \cline {2 -3}
							& Followed by mo		& Is it followed by an $<$mo$>$ node 		\\ \cline {2 -3}
							& $\&\#8289;$			& Is it followed by a Function Application 			\\ \cline {2 -3}
							& Parent's name		 	& The name of its parent node 							\\ \cline {2 -3}
							& Name		 			& The name of the identifier 							\\ \hline
Text feature				& Category 				& Relation between category name and candidate translation	\\ \cline {2 -3}
							& Unigram 				& Vector represents unigram feature		  	\\ \cline {2 -3}
							& Bigram 				& Vector represents bigram feature	 	\\ \cline {2 -3}
							& Trigram 				& Vector represents trigram feature		 	\\ \hline
\multicolumn{2}{|l|}{Candidate translation}			& One of $n$ candidate translations of the $mi$ element 		 	\\ \hline
\end{tabular}
\label{tab:fea}
\end{table}

% MathML feature
There were six Presentation MathML features in our experiment.
The first one determines whether the $mi$ element is the only child of its parent.
The relation between the $mi$ element and its surrounding $mo$ elements is encoded in the following three features.
The last two features represent the name of the $mi$ element and its parent.
Among these features, the name of the $mi$ element is the most important feature.

% Description
Among the text features, the first one is the category that mathematical expression belongs to.
In mathematical resource websites, such as the Wolfram Functions Site, mathematical expressions belong to different categories.
But usually we do not have the text surrounding these mathematical expressions.
We then can calculate the relation between the category name and the Content translation of each $mi$ element.
The relation has one of three values: the same as the Content translation, contains the Content translation, or does not contain the Content translation.

In case we have the text surrounding or the description of the mathematical expressions, we can use n-gram features~\cite{Ngram}.
In this paper, we use unigram, bigram and trigram features.
These features are implemented as the vectors containing the n-grams which appear in the training data.
We will assign each instance into one of two classes, depending on the candidate translation.
The class is `true' if the candidate translation is the correct Content translation of the $mi$ element, and `false' otherwise.

% Plugin to SMT system
\subsection{Translation}
After disambiguation, we use the result to enhance the semantic enrichment of a statistical-machine-translation-based system.
The input of this module includes a Presentation MathML expression, a set of rules for translation, and the output from the disambiguation module. The output of this module is the Content MathML expression which represents the meaning of the Presentation MathML expression.
If there is only one mapping from a Presentation element, that Content element is chosen.
If the disambiguation module accepts more than two mappings from a Presentation element, the Content element with higher probability is chosen.

%% file: 4-eva.tex
\section{Evaluation}

% Evaluation Setup
The first dataset for the experiments is the Wolfram Functions site~\cite{Wolfram}.
This site was created as a resource for educational, mathematical, and scientific communities.
All formulas on this site are available in both Presentation MathML and Content MathML format.
The only text information on this dataset is the function category of each mathematical expression.
In our experiments, we used 136,685 mathematical expressions divided into seven categories.

The second dataset for the experiments is the Archives of the Association for Computational Linguistics Corpus~\cite{ACL-ARC} (ACL-ARC).
It contains mathematical expressions extracted from scientific papers in the area of Computational Linguistics and Language Technology.
Currently, we use mathematical expressions drawn from 20 papers which were selected from this dataset.
We have manually annotated all mathematical expressions with MathML parallel Markup and their textual descriptions.
Out of 2,065 mathematical expressions in the dataset, only 648 expressions have their own description.
Table \ref{tab:aclex} shows examples of mathematical expressions and their description in ACL-ARC dataset.

\begin{table}[htb]
\centering
\caption{Examples of mathematical expressions and their description in ACL-ARC dataset}
\begin{tabular}{|L{4cm}|L{4cm}|L{3.8cm}|}
\hline
\centering\textbf{Textual description} & \centering\textbf{MathML Presentation expression} & \centering\textbf{MathML Content expressions} \tabularnewline \hline
a word to be translated & $<$mrow$>$ $<$mi$>$w$<$/mi$>$ $<$/mrow$>$ & $<$ci$>$w$<$/ci$>$\\ \hline
a word in a dependency relationship & $<$mrow$>$ $<$mi$>$w$<$/mi$>$ $<$/mrow$>$ & $<$ci$>$w$<$/ci$>$\\ \hline
a matrix & $<$mrow$>$ $<$mi$>$t$<$/mi$>$ $<$/mrow$>$ & $<$ci$>$t$<$/ci$>$\\ \hline
a similarity matrix which specifies the similarity between individual elements & $<$mrow$>$ $<$mi$>$sim$<$/mi$>$ $<$/mrow$>$ & $<$ci$>$sim$<$/ci$>$\\ \hline
argument & $<$mrow$>$ $<$msub$>$ $<$mi$>$S$<$/mi$>$ $<$msub$>$ $<$mi$>$j$<$/mi$>$ $<$mi$>$i$<$/mi$>$ $<$/msub$>$ $<$/msub$>$ $<$/mrow$>$ & $<$apply$>$ $<$selector /$>$ $<$ci$>$S$<$/ci$>$ $<$apply$>$ $<$selector /$>$ $<$ci$>$j$<$/ci$>$ $<$ci$>$i$<$/ci$>$ $<$/apply$>$ $<$/apply$>$\\ \hline
The LM probabilities & $<$mrow$>$ $<$mi$>$P$<$/mi$>$ $<$mo$>$⁡$<$/mo$>$ $<$mrow$>$ $<$mo$>$($<$/mo$>$ $<$mrow$>$ $<$mi$>$v$<$/mi$>$ $<$mo$>\vert<$/mo$>$ $<$mrow$>$ $<$mi$>$Parent$<$/mi$>$ $<$mo$>$⁡$<$/mo$>$ $<$mrow$>$ $<$mo$>$($<$/mo$>$ $<$mi$>$v$<$/mi$>$ $<$mo$>$)$<$/mo$>$ $<$/mrow$>$ $<$/mrow$>$ $<$/mrow$>$ $<$mo$>$)$<$/mo$>$ $<$/mrow$>$ $<$/mrow$>$ & $<$apply$>$ $<$ci$>$P$<$/ci$>$ $<$apply$>$ $<$ci$>\vert<$/ci$>$ $<$ci$>$v$<$/ci$>$ $<$apply$>$ $<$ci$>$Parent$<$/ci$>$ $<$ci$>$v$<$/ci$>$ $<$/apply$>$ $<$/apply$>$ $<$/apply$>$\\ \hline
\end{tabular}
\label{tab:aclex}
\end{table}

The evaluation was done using two metrics: accuracy score for disambiguation and tree edit distance rate score for semantic enrichment.
The accuracy score of disambiguation is the ratio of correctly classified instances to total instances.
The tree edit distance rate (TEDR) score~\cite{JSAI} is defined as the ratio of (1) the minimal cost of transforming the generated into the reference Content MathML tree using edit operations and (2) the maximum number of nodes of the generated and the reference Content MathML tree.
We also compare our semantic enrichment results to the results of Nghiem et al.

% Disambiguation results
First, we set up an experiment to examine the disambiguation result on each Presentation MathML $mi$ element.
In this experiment, we compare three systems.
The first system uses both Presentation MathML and text features.
The second system uses only Presentation MathML features.
The last system chooses the interpretation with highest probability.

Training and testing were performed using ten-fold cross-validation.
For each category, we partitioned the original corpus into ten subsets.
Of the ten subsets, we retained a single subset as validation data for testing the model, remaining subsets are used as training data.
The cross-validation process was repeated ten times, and the ten results from the folds then averaged to produce a single estimate.
Table \ref{tab:disres} shows the results of the disambiguation component.

\begin{table}[htb]
\centering
\caption{Disambiguation accuracy}
\begin{tabular}{|L{3.7cm}|R{1.6cm}|R{1.6cm}|R{1.6cm}|R{1.6cm}|}
\hline
\centering\textbf{Category} 	& \centering\textbf{Number of instances} & \centering\textbf{With text features} & \centering\textbf{Without text features} & \centering\textbf{Most frequent} \tabularnewline \hline
ACL-ARC		 		& 2,996 	& 92.9573 & \textbf{93.7583} & 93.4246 \\ \hline
Bessel-TypeFunctions& 1,352 	& \textbf{92.8254} & 92.3077 & 86.0947\\ \hline
Constants			& 714 		& \textbf{91.1765} & 90.3361 & 83.7535\\ \hline
ElementaryFunctions & 6,073 	& 96.1963 & \textbf{96.3774} & 89.6427\\ \hline
GammaBetaErf		& 3,816 	& \textbf{95.2830} & 94.4706 & 78.0136 \\ \hline
HypergeometricFunctions& 72,006 & \textbf{97.5571} & 97.0697 & 88.0746\\ \hline
IntegerFunctions	& 11,955 	& \textbf{95.8009} & 95.1652 & 90.0711 \\ \hline
Polynomials			& 5,905 	& \textbf{98.2388} & 95.3091 & 87.3328 \\ \hline
All WFS Data		& 320,726 	& \textbf{98.9243} & 98.4398 & 92.7025 \\ \hline
\end{tabular}
\label{tab:disres}
\end{table}

The results in Table \ref{tab:disres} show that disambiguation result using SVM outperformed the `most frequent' method.
The reason `most frequent' method got high scores is because mathematical elements often have a preferred meaning.
The systems that used only Presentation MathML features achieved even better scores, because they use surrounding mathematical elements.
It is interesting to note that on the ACL-ARC data, the `most frequent' system get higher score than the system with text features.
Overall, on WFS data, we gained 5 to 16 percent accuracy improvements. 

The systems that also used text features outperform the systems that used only Presentation MathML features in most of WFS categories.
This result may be explained by the fact that the category of a mathematical expression is closely related to that expression.
Contrary to expectations, this study did not find any improvement in ACL-ARC data.
It seems possible that these results are due to the lack of training data and the sparseness of n-gram features.
This finding was unexpected and suggests that in order to use n-gram text features, we need more data.

Second, we set up an experiment to examine the semantic enrichment result.
The results from disambiguation component are used in the semantic enrichment system.
We compare three systems: with text feature, without text feature, and the system of Nghiem et al. which used `most frequent' method.
In this experiment, we use 90 percent of expressions for training both SVM-based disambiguation and translation components.
We use the other 10 percent of expressions for testing.
Table \ref{tab:tranres} shows the translation result.

\begin{table}[htb]
\centering
\caption{Semantic enrichment TEDR}
\begin{tabular}{|L{3.7cm}|R{1.6cm}|R{1.6cm}|R{1.6cm}|R{1.6cm}|}
\hline
\centering\textbf{Category} & \centering\textbf{Number of expression} & \centering\textbf{With text feature} & \centering\textbf{Without text feature} & \centering\textbf{Most frequent} \tabularnewline \hline
Bessel-TypeFunctions& 701 & \textbf{18.0604} & \textbf{18.0604} & 18.4118 \\ \hline
Constants			& 555 & \textbf{33.9016} & 34.0328 & 34.6230 \\ \hline
ElementaryFunctions & 9,537 & 7.4879 & \textbf{7.4809} & 7.7343 \\ \hline
GammaBetaErf		& 1,558 & \textbf{17.2308} & 17.2851 & 18.4796 \\ \hline
HypergeometricFunctions& 9,347 & \textbf{49.4678} & 49.4797 & 49.6902 \\ \hline
IntegerFunctions	& 1,175 & \textbf{20.5292} & 20.5874 & 20.9945 \\ \hline
Polynomials			& 727 & \textbf{19.6309} & 19.7987 & 20.2685 \\ \hline
All WFS Data		& 23,600 & \textbf{29.0707} & 29.0869 & 29.2769 \\ \hline
\end{tabular}
\label{tab:tranres}
\end{table}

The results in Table \ref{tab:tranres} show that combining disambiguation and statistical machine translation improved the system.
Expressions in `Gamma Beta Erf' category benefit from the disambiguation module the most with 1.2 percent error rate reduction.
Less ambiguity in elementary functions might lead to lower performance in `Elementary Functions' category.
We did not evaluate on ACL-ARC data because the disambiguation result was almost the same as the `most frequent' method.
Overall, on WFS data, we achieved 0.2 to 1.2 percent error rate reduction.

%% file: 5-con.tex
\section{Conclusion}

In this paper, we have presented a new approach to the semantic enrichment for mathematical expression problem.
Our approach, which combines statistical machine translation and disambiguation component, shows promise.
This study has shown that the disambiguation component using presentation features improved the system performance.
The use of text features, especially the category of each expression, also played an important role in the disambiguation of mathematical elements.
Experimental results of this study showed that our system achieves improvements over prior systems.

% Future research
This research has raised many questions in need of further investigation.
One question is finding and combining new features, such as the style of the font, for the disambiguation task.
Another possible improvement is making use of co-occurrence of mathematical elements in the same document.
In the scope of this paper, we only disambiguated lexical ambiguities of mathematical expressions.
Structural ambiguities should also be considered to achieve better results.
The evidence from this study suggests that in a small dataset, descriptions of mathematical expressions did not improve the system performance.
Further work needs to be done to establish whether descriptions of mathematical expressions contribute to the the task in a larger dataset.